\magnification 1200

%
%
\newdimen\FigSize       \FigSize=.9\hsize 
%
\newskip\abovefigskip   \newskip\belowfigskip
\gdef\epsfig#1;#2;{\par\vskip\abovefigskip\penalty -500
   {\everypar={}\epsfxsize=#1\nd
    \centerline{\epsfbox{#2}}}%
    \vskip\belowfigskip}%
%
\newskip\figtitleskip
\gdef\tepsfig#1;#2;#3{\par\vskip\abovefigskip\penalty -500
   {\everypar={}\epsfxsize=#1\nd
    \vbox
      {\centerline{\epsfbox{#2}}\vskip\figtitleskip
       \centerline{\figtitlefont#3}}}%
    \vskip\belowfigskip}%
%
\newcount\FigNr \global\FigNr=0
\gdef\nepsfig#1;#2;#3{\global\advance\FigNr by 1
   \tepsfig#1;#2;{Figure\space\the\FigNr.\space#3}}%
%
%
%
\gdef\ipsfig#1;#2;{
   \midinsert{\everypar={}\epsfxsize=#1\nd
              \centerline{\epsfbox{#2}}}%
   \endinsert}%
%
\gdef\tipsfig#1;#2;#3{\midinsert
   {\everypar={}\epsfxsize=#1\nd
    \vbox{\centerline{\epsfbox{#2}}%
          \vskip\figtitleskip
          \centerline{\figtitlefont#3}}}\endinsert}%
%
\gdef\nipsfig#1;#2;#3{\global\advance\FigNr by1%
  \tipsfig#1;#2;{Figure\space\the\FigNr.\space#3}}%
\newread\epsffilein    
\newif\ifepsffileok    
\newif\ifepsfbbfound   
\newif\ifepsfverbose   
\newdimen\epsfxsize    
\newdimen\epsfysize    
\newdimen\epsftsize    
\newdimen\epsfrsize    
\newdimen\epsftmp      
\newdimen\pspoints     
\pspoints=1bp          
\epsfxsize=0pt         
\epsfysize=0pt         
\def\epsfbox#1{\global\def\epsfllx{72}\global\def\epsflly{72}%
   \global\def\epsfurx{540}\global\def\epsfury{720}%
   \def\lbracket{[}\def\testit{#1}\ifx\testit\lbracket
   \let\next=\epsfgetlitbb\else\let\next=\epsfnormal\fi\next{#1}}%
\def\epsfgetlitbb#1#2 #3 #4 #5]#6{\epsfgrab #2 #3 #4 #5 .\\%
   \epsfsetgraph{#6}}%
\def\epsfnormal#1{\epsfgetbb{#1}\epsfsetgraph{#1}}%
\def\epsfgetbb#1{%
%
%
\openin\epsffilein=#1
\ifeof\epsffilein\errmessage{I couldn't open #1, will ignore it}\else
%
%
   {\epsffileoktrue \chardef\other=12
    \def\do##1{\catcode`##1=\other}\dospecials \catcode`\ =10
    \loop
       \read\epsffilein to \epsffileline
       \ifeof\epsffilein\epsffileokfalse\else
%
%
          \expandafter\epsfaux\epsffileline:. \\%
       \fi
   \ifepsffileok\repeat
   \ifepsfbbfound\else
    \ifepsfverbose\message{No bounding box comment in #1; using
defaults}\fi\fi
   }\closein\epsffilein\fi}%
%
%
\def\epsfsetgraph#1{%
   \epsfrsize=\epsfury\pspoints
   \advance\epsfrsize by-\epsflly\pspoints
   \epsftsize=\epsfurx\pspoints
   \advance\epsftsize by-\epsfllx\pspoints
%
%
   \epsfxsize\epsfsize\epsftsize\epsfrsize
   \ifnum\epsfxsize=0 \ifnum\epsfysize=0
      \epsfxsize=\epsftsize \epsfysize=\epsfrsize
%
arithmetic!
%
     \else\epsftmp=\epsftsize \divide\epsftmp\epsfrsize
       \epsfxsize=\epsfysize \multiply\epsfxsize\epsftmp
       \multiply\epsftmp\epsfrsize \advance\epsftsize-\epsftmp
       \epsftmp=\epsfysize
       \loop \advance\epsftsize\epsftsize \divide\epsftmp 2
       \ifnum\epsftmp>0
          \ifnum\epsftsize<\epsfrsize\else
             \advance\epsftsize-\epsfrsize \advance\epsfxsize\epsftmp
\fi
       \repeat
     \fi
   \else\epsftmp=\epsfrsize \divide\epsftmp\epsftsize
     \epsfysize=\epsfxsize \multiply\epsfysize\epsftmp
     \multiply\epsftmp\epsftsize \advance\epsfrsize-\epsftmp
     \epsftmp=\epsfxsize
     \loop \advance\epsfrsize\epsfrsize \divide\epsftmp 2
     \ifnum\epsftmp>0
        \ifnum\epsfrsize<\epsftsize\else
           \advance\epsfrsize-\epsftsize \advance\epsfysize\epsftmp \fi
     \repeat
   \fi
%
%
   \ifepsfverbose\message{#1: width=\the\epsfxsize,
height=\the\epsfysize}\fi
   \epsftmp=10\epsfxsize \divide\epsftmp\pspoints
   \vbox to\epsfysize{\vfil\hbox to\epsfxsize{%
      \includegraphics{#1}%
      \hfil}}%
\epsfxsize=0pt\epsfysize=0pt}%
%
%
{\catcode`\%=12
\global\let\epsfpercent=
%
%
\long\def\epsfaux#1#2:#3\\{\ifx#1\epsfpercent
   \def\testit{#2}\ifx\testit\epsfbblit
      \epsfgrab #3 . . . \\%
      \epsffileokfalse
      \global\epsfbbfoundtrue
   \fi\else\ifx#1\par\else\epsffileokfalse\fi\fi}%
%
%
\def\epsfgrab #1 #2 #3 #4 #5\\{%
   \global\def\epsfllx{#1}\ifx\epsfllx\empty
      \epsfgrab #2 #3 #4 #5 .\\\else
   \global\def\epsflly{#2}%
   \global\def\epsfurx{#3}\global\def\epsfury{#4}\fi}%
%
%
\def\epsfsize#1#2{\epsfxsize}%
%
%

\epsfverbosetrue                        
\abovefigskip=\baselineskip             
\belowfigskip=\baselineskip             
\global\let\figtitlefont\bf             
\global\figtitleskip=.5\baselineskip    

\font\tenmsb=msbm10   
\font\sevenmsb=msbm7
\font\fivemsb=msbm5
\newfam\msbfam
\textfont\msbfam=\tenmsb
\scriptfont\msbfam=\sevenmsb
\scriptscriptfont\msbfam=\fivemsb
\def\Bbb#1{\fam\msbfam\relax#1}
\let\nd\noindent 
\def\qed{\hbox{\hskip 6pt\vrule width6pt height7pt depth1pt \hskip1pt}}
\def\natural{{\rm I\kern-.18em N}}
\newskip\ttglue


\def\eightpoint{\def\rm{\fam0\eightrm}  
  \textfont0=\eightrm \scriptfont0=\sixrm \scriptscriptfont0=\fiverm
  \textfont1=\eighti  \scriptfont1=\sixi  \scriptscriptfont1=\fivei
  \textfont2=\eightsy  \scriptfont2=\sixsy  \scriptscriptfont2=\fivesy
  \textfont3=\tenex  \scriptfont3=\tenex  \scriptscriptfont3=\tenex
  \textfont\itfam=\eightit  \def\it{\fam\itfam\eightit}
  \textfont\slfam=\eightsl  \def\sl{\fam\slfam\eightsl}
  \textfont\ttfam=\eighttt  \def\tt{\fam\ttfam\eighttt}
  \textfont\bffam=\eightbf  \scriptfont\bffam=\sixbf
    \scriptscriptfont\bffam=\fivebf  \def\bf{\fam\bffam\eightbf}
  \tt  \ttglue=.5em plus.25em minus.15em
  \normalbaselineskip=9pt
  \setbox\strutbox=\hbox{\vrule height7pt depth2pt width0pt}
  \let\sc=\sixrm  \let\big=\eightbig \normalbaselines\rm}

\font\eightrm=cmr8 \font\sixrm=cmr6 \font\fiverm=cmr5
\font\eighti=cmmi8  \font\sixi=cmmi6   \font\fivei=cmmi5
\font\eightsy=cmsy8  \font\sixsy=cmsy6 \font\fivesy=cmsy5
\font\eightit=cmti8  \font\eightsl=cmsl8  \font\eighttt=cmtt8
\font\eightbf=cmbx8  \font\sixbf=cmbx6 \font\fivebf=cmbx5

\def\eightbig#1{{\hbox{$\textfont0=\ninerm\textfont2=\ninesy
        \left#1\vbox to6.5pt{}\right.\enspace$}}}

\def\ninepoint{\def\rm{\fam0\ninerm}  
  \textfont0=\ninerm \scriptfont0=\sixrm \scriptscriptfont0=\fiverm
  \textfont1=\ninei  \scriptfont1=\sixi  \scriptscriptfont1=\fivei
  \textfont2=\ninesy  \scriptfont2=\sixsy  \scriptscriptfont2=\fivesy
  \textfont3=\tenex  \scriptfont3=\tenex  \scriptscriptfont3=\tenex
  \textfont\itfam=\nineit  \def\it{\fam\itfam\nineit}
  \textfont\slfam=\ninesl  \def\sl{\fam\slfam\ninesl}
  \textfont\ttfam=\ninett  \def\tt{\fam\ttfam\ninett}
  \textfont\bffam=\ninebf  \scriptfont\bffam=\sixbf
    \scriptscriptfont\bffam=\fivebf  \def\bf{\fam\bffam\ninebf}
  \tt  \ttglue=.5em plus.25em minus.15em
  \normalbaselineskip=11pt
  \setbox\strutbox=\hbox{\vrule height8pt depth3pt width0pt}
  \let\sc=\sevenrm  \let\big=\ninebig \normalbaselines\rm}

\font\ninerm=cmr9 \font\sixrm=cmr6 \font\fiverm=cmr5
\font\ninei=cmmi9  \font\sixi=cmmi6   \font\fivei=cmmi5
\font\ninesy=cmsy9  \font\sixsy=cmsy6 \font\fivesy=cmsy5
\font\nineit=cmti9  \font\ninesl=cmsl9  \font\ninett=cmtt9
\font\ninebf=cmbx9  \font\sixbf=cmbx6 \font\fivebf=cmbx5
\def\ninebig#1{{\hbox{$\textfont0=\tenrm\textfont2=\tensy
        \left#1\vbox to7.25pt{}\right.$}}}

\def\R{{\Bbb R}}
\def\Z{{\Bbb Z}}
\def\chix{{\raise.5ex\hbox{$\chi$}}}
\def\chixa{{\chix\lower.2em\hbox{$_A$}}}

\def\real{{\rm I\kern-.2em R}}
\def\integer{{\rm Z\kern-.32em Z}}
\def\complex{\kern.1em{\raise.47ex\hbox{
            $\scriptscriptstyle |$}}\kern-.40em{\rm C}}
\def\vs#1 {\vskip#1truein}
\def\hs#1 {\hskip#1truein}
  \hsize=6.2truein \hoffset=.23truein 
  \vsize=8.8truein 
\pageno=1 \baselineskip=12pt
  \parskip=0 pt \parindent=20pt 
\overfullrule=0pt \lineskip=0pt \lineskiplimit=0pt
  \hbadness=10000 \vbadness=10000 
     \pageno=0
     
     \footline{\ifnum\pageno=0\hss\else\hss\tenrm\folio\hss\fi}
     \hbox{}
     \vskip 1truein\centerline{{\bf Random Close Packing in a Granular Model}}
     \vskip .2truein\centerline{by}
     \vskip .2truein
\centerline{{David Aristoff}
\ \ and\ \  {Charles Radin}
\footnote{*}{Research supported in part by NSF Grant DMS-0700120\hfil}}

\vskip .1truein
\centerline{ Mathematics Department, University of Texas, Austin, TX 78712} 
\vs.5 \centerline{{\bf Abstract}} 

\vs.1 \nd We introduce a 2-dimensional lattice model of
     granular matter. We use a combination of proof and simulation to
     demonstrate an order/disorder phase transition in the model, to
     which we associate the granular phenomenon of random close
     packing.

\vs3
\centerline{September, 2009}
\vs1
\centerline{PACS Classification:\ \ 45.70.Cc, 81.05.Rm, 05.70.Ce}
     \vfill\eject

\nd {\bf 0. Introduction.} \vs.1 

Granular materials, such as a static pile of sand or salt grains
sedimented in a fluid such as air, exhibit interesting characteristic
behavior at certain volume fractions. For sand in air the lowest
possible volume fraction (called the {\it random loose packing density}) is
about 0.58, and the highest possible volume fraction is about 0.74. 
In other words a sand pile can exist with volume fraction
anywhere in the interval $(0.58,\ 0.74)$. Within this range there are
also: the {\it critical state density}, about 0.60, and the {\it
random close packing density}, about 0.64 [dG]. In this paper we
consider a toy model for granular materials, the goal being to model
granular behavior near the random close packing density. Our results
support the interpretation in [Ra] of the phenomenon of random close
packing as an {\it order/disorder phase transition}; we show in our
model that at high density the system is sensitive to the boundary
conditions while at low density it is not, with a perfectly sharp
transition in between.

Our model is 2-dimensional and consists of nonoverlapping, parallel,
hexagonal ``grains'' for which the centers (and corners) lie on sites
of the planar triangular lattice: $\{m(1/2,\sqrt{3}/2) +
n(1,0)\,|\,m,n\in \Z\}$ (see Figure 1). 
\vs-.1 \nd
\epsfig .6\hsize;  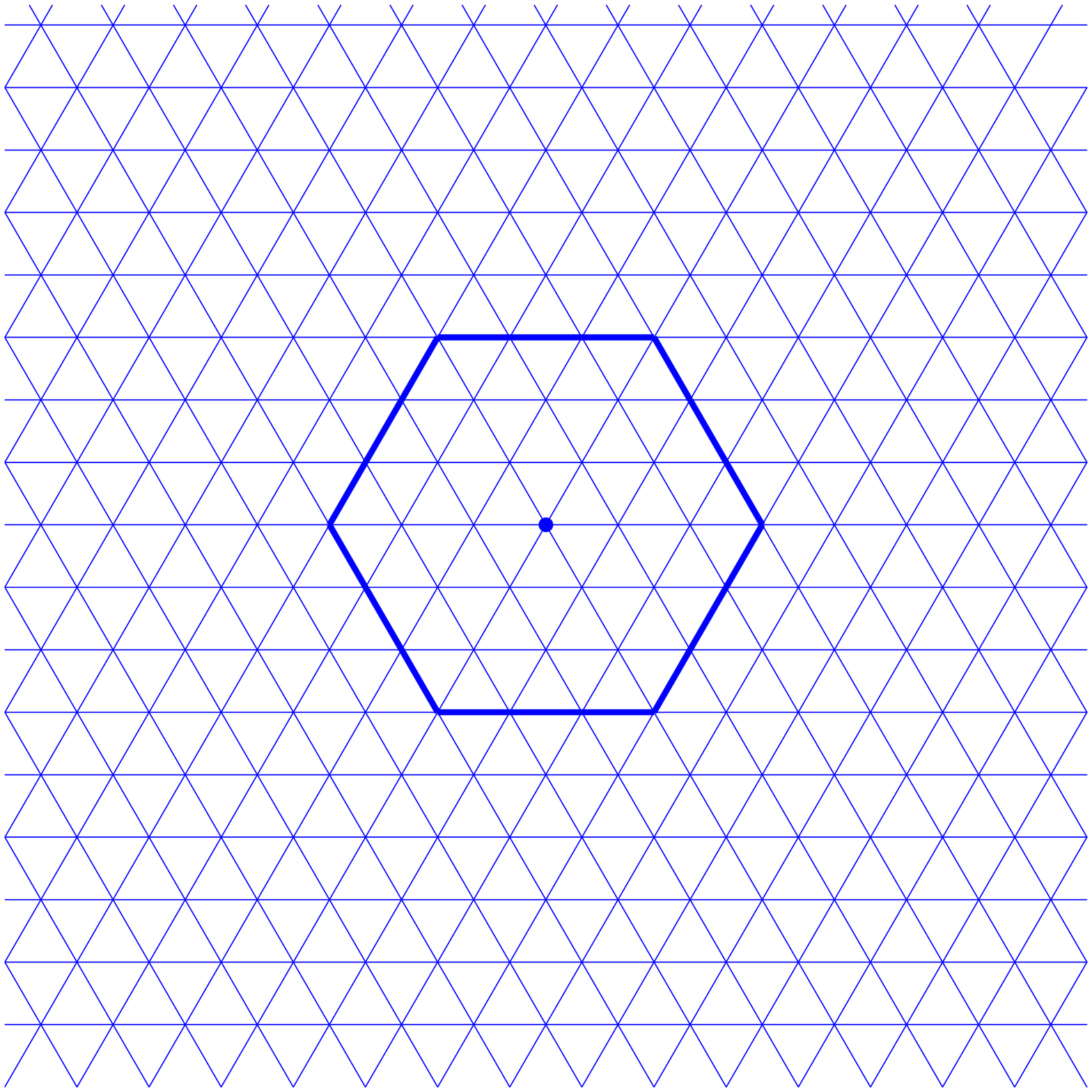;
\vs-.1 \nd
\centerline{Figure 1. An hexagonal grain on the triangular lattice.}
\vs.2

To account for the effects of gravity and
friction we impose the condition that a configuration is {allowed} or
{\it legal} only if each hexagonal grain intersects 
one of the three upper edges of another hexagonal grain, such that 
the latter grain has a center below that of the former (see Figure
2). Nearest neighbor sites in the lattice have separation 1, and the
hexagons all have the same integral side length $s$. For the
simulations described below we use $s=2,3$ or 4; for our proofs any $s\ge 1$
suffices.

We use a ``grand canonical'' version of the Edwards model [EO] of
granular matter; in this version the probability $Pr_{_V}(A)$ of a
legal configuration $A$ of $n$ particles in a fixed volume $V$ is
$e^{\mu n}/Z_{_V}(\mu)$, where $\mu\in \R$ is a parameter and
$Z_{_V}(\mu)$ is the normalization constant (grand partition
function); the ``infinite volume limit'' [Ru] is then taken, in which
$V\to \R^2$.

Note that this model is a variation on the hard-core lattice-gas
models of classical statistical mechanics, introduced by Lee and Yang
in [LY], which use Peierls contours to prove a phase transition. (See
[Gi, HP] for some later developments.) Specifically, in this method and
for ``extended'' hard-cores in which particles are larger than a
single lattice site, one proves that at all sufficiently high values
of $\mu$ the model exhibits long range positional order, being
sensitive at the middle of configurations to the precise relative location of
the distant boundary, while at all sufficiently low values of $\mu$
the model is (easily) shown to behave as a dilute, disordered fluid,
insensitive to the boundary.

\vs-.1 \nd
\epsfig .8\hsize;  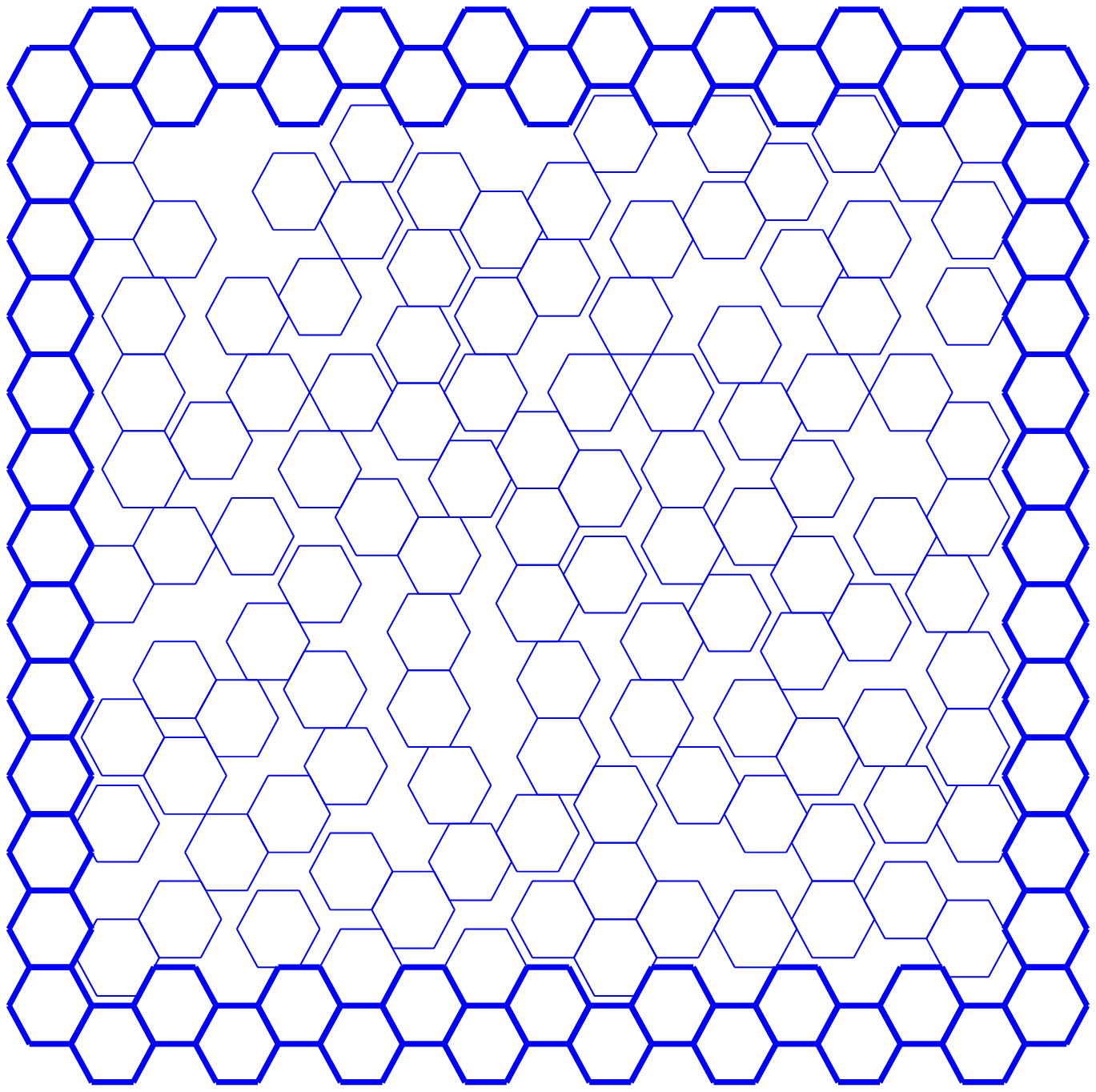;
\vs-.1 \nd
\centerline{Figure 2. A legal configuration (boundary hexagons are in boldface).}
\vs.2
For our granular model we are able to prove long range positional
order for all sufficiently high values of $\mu$, but not disordered
behavior at low $\mu$. In place of a proof for the latter we have
performed Monte Carlo simulations at low values of $\mu$ to
demonstrate disorder. The proof for high $\mu$ is given in Section 1
and the numerical results for low $\mu$ are in Section 2.

We note that there was a previous granular adaption of the old hard
hexagon models by Monasson and Pouliquen [MP]. Their model differs in
several important details, for instance their use of periodic boundary
conditions; more
important is that they employ their model in a study of
entropy rather than random close packing.

\vs.3
\nd 
{\bf 1. Proof for high $\mu$.}
\vs.2

Consider a regular triangular lattice with distance between nearest neighbor 
lattice sites equal to $1$. We consider configurations of hard-core 
parallel regular {\it hexagons}, where a hexagon is
centered at a lattice point and has side length equal to a fixed 
integer $s \ge 1$. The hexagons are all inside a square container 
$V$ whose boundary consists of hexagons which intersect in full edges (see Figure 2).

We call a configuration of nonoverlapping hexagons inside the
boundary {\it legal} if each hexagon $h$ intersects one of the three 
upper edges of another hexagon $h'$. (We also require that $h'$ is centered 
strictly below $h$.) We call $h'$ a {\it support} of $h$. We let the number of
hexagons inside the boundary (called {\it interior hexagons}), $n$,
vary, and fix $\mu \ge 0$. The probability of seeing a given 
configuration $A$ is $Pr(A)= e^{\mu n}/Z$, where $n$ is
the number of interior hexagons in $A$ and $Z=Z(\mu)$ is the
normalization. (For simplicity the notation will ignore dependence on the
container $V$.)

Two hexagons $h$, $h'$ are said to be {\it linked} if their intersection 
is a full edge (i.e. a line segment of length $s$), or if there is a 
sequence of hexagons $h_0 = h, h_1,..., h_m = h'$ such that $h_i$ intersects 
$h_{i+1}$ in a full edge for $i=0,...,m-1$. In particular, 
the hexagons on the boundary are 
all linked. We are interested in the event that the 
origin lies inside a hexagon linked to a boundary
hexagon; we call this event $0_{LB}$.

A {\it triangle} is a closed regular triangle with side
length $1$ and vertices at lattice sites. 
Given a configuration $A$ inside $V$, we define a {\it contour} 
in $A$ to be one of the connected components of
the union of all triangles in $V$ not covered 
by hexagons in $A$, and all line segments in $V$ of length
strictly less than $s$ which are intersections of neighboring
hexagons in $A$. An {\it outer contour} is a contour which intersects a boundary 
hexagon or a hexagon linked to the boundary (see Figure 3). Note that the topological boundary 
of a contour $C$ contains a closed curve $\gamma$ which encloses an area containing 
the entire contour. We call the region enclosed by $\gamma$ the {\it region
enclosed by $C$}.

A {\it sublattice} is a set of points which are the centers 
of a collection of hexagons which tile the plane. There are 
$3s^2$ distinct sublattices. Note that any set 
of hexagons which are linked corresponds to a single sublattice; in 
particular the boundary hexagons define a sublattice which we call 
the {\it boundary sublattice}. We
say that a hexagon is {\it on the boundary sublattice} if its center is 
in the sublattice defined by the boundary hexagons.

\vs.3 \nd
{\bf Definition 1}.
Consider a hexagon $h$ of side length $s$ centered at the origin. Let $S$ be a set of 
$3s^2$ lattice sites in $h$ such that $S$ has exactly one representative of each sublattice. 
We are interested in the event that there is a hexagon centered in $S$ which is on the 
boundary sublattice; we call this event $O_B$.
\vs.05
\vs-.1 \nd
\epsfig .7\hsize;  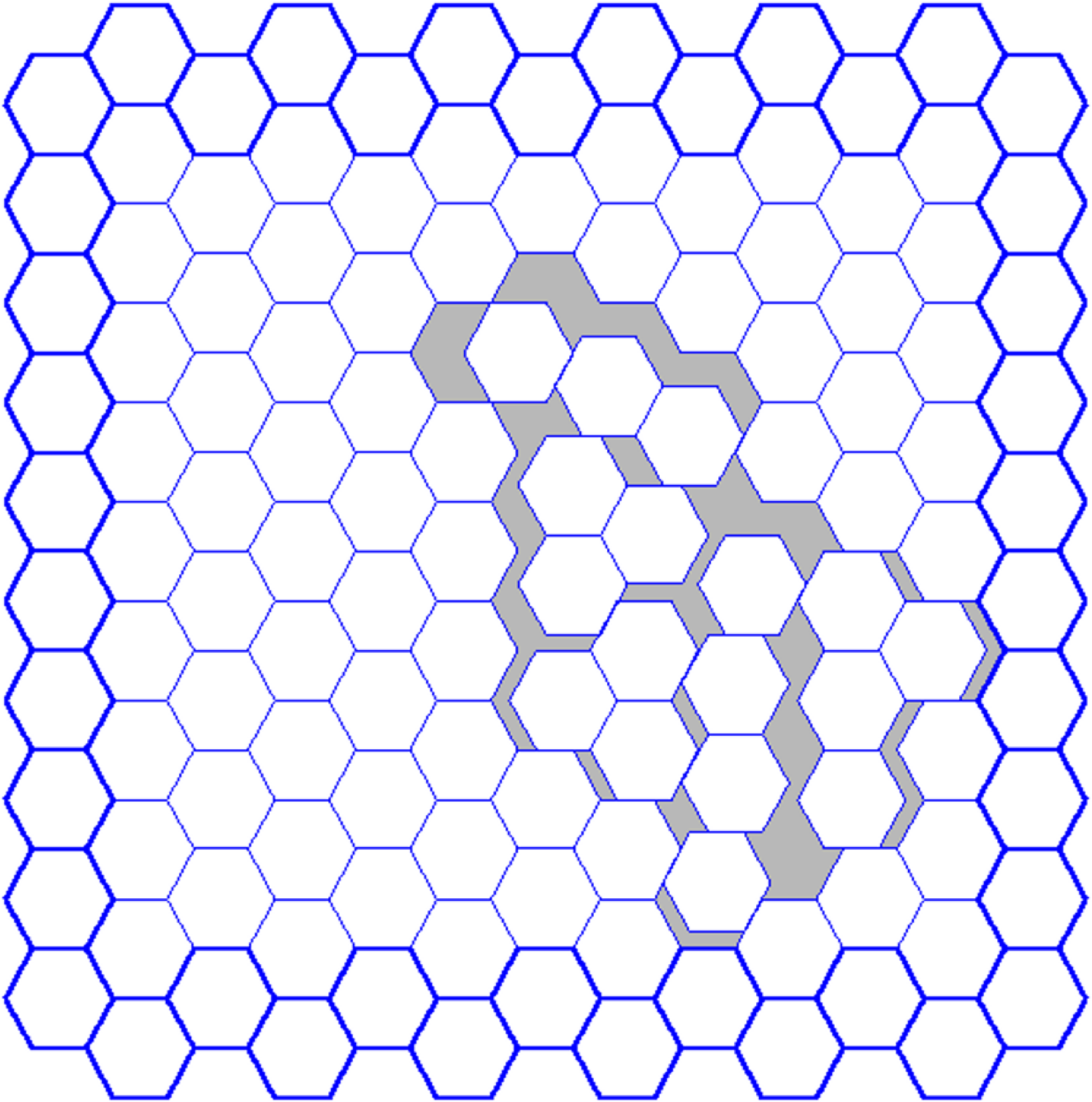;
\vs-.1 \nd
\centerline{Figure 3. An outer contour (shaded region).}
\vs.2

\vs.2 \nd 
{\bf Lemma 1.} If there is a hexagon centered in $S$ which is not on the boundary 
sublattice, then this hexagon is not linked to the boundary.
\vs.1 \nd
{\bf Proof.} If a hexagon centered in $S$ 
is linked to the boundary, then it is on the sublattice defined by the boundary, 
since a set of linked hexagons corresponds to a single sublattice. \qed

\vs.2 \nd 
{\bf Lemma 2.} If there is no hexagon centered in $S$, or if there is 
a hexagon centered in $S$ not linked to the boundary, then there is an outer 
contour $C$ such that the origin is in the region enclosed by $C$.
\vs.1 \nd 
{\bf Proof.} If there is no hexagon centered in $S$, then the origin 
itself is inside a contour and the result follows. Now assume there 
is a hexagon centered in $S$, and consider the contrapositive. If there 
is no outer contour enclosing the origin, then there is no contour at 
all enclosing the origin. Thus the hexagon centered in $S$ is linked 
to the boundary. \qed

\vs.2 \nd 
{\bf Corollary 1.} If there is no hexagon centered in $S$, or if there 
is a hexagon centered in $S$ which is not on the boundary sublattice,
 then there is an outer contour enclosing the origin.
\vs.1 \nd 
{\bf Proof.} This follows from Lemmas 1 and 2. \qed

\vs.2
Let $[0_{B}]^c$ be the complement of the event $0_{B}$. We will give an upper 
bound for $P([0_{B}]^c)$ by using a Peierls-type argument. We will first show 
that the probability of seeing a fixed contour is exponentially small
for large $\mu$. Then we will use a counting argument to get
an upper bound on the number of possible contours.

The {\it size} of a contour is defined as its area in units of the
area of a hexagon. We will see shortly that the size of a contour must
be an integer. Given a contour $C$, let $E$ be the region enclosed by $C$. 
The closures of the connected 
components of the complement of $C \cap E$ will be called 
{\it $C$-interior regions}. Note that the
hexagons with edges on the topological boundary of a $C$-interior 
region $R$ must be linked; we say such hexagons are on the {\it outside} of the $R$, 
and we say the remaining hexagons in $R$ are on the {\it inside} of $R$.

We say that we {\it shift} a $C$-interior region $R$ if we translate
$R$ while holding all other hexagons fixed. The translation must be 
given by a difference $x-y$ where $x$ and $y$ are lattice sites.
Note that the relative positions of hexagons in a $C$-interior region 
$R$ are unchanged by a shift. 

Given a contour $C$, we say a $C$-interior region $R$ is {\it on the boundary 
sublattice} if the hexagons on the outside of $R$ are on the boundary sublattice.

\vs.2 \nd 
{\bf Lemma 3.} For any outer contour $C$ there is a sequence 
of shifts of all of the $C$-interior regions such that in the resulting 
configuration, each of the shifted interior regions is on the boundary 
sublattice and no hexagons overlap one another.

\vs.1 \nd 
{\bf Proof.} For each site $x$ on the boundary sublattice define a 
neighborhood $N_x$ of $x$ as follows. Let $h$ be a regular hexagon of 
side length $s$ centered at $x$. Then $N_x$ consists of all lattice sites 
in $h$ except those on any of the bottom three edges of $h$.

Note that the neighborhoods $N_x$ are disjoint and together cover all the 
lattice sites. Create a sequence of shifts of the $C$-interior regions 
as follows. For each $C$-interior region $R$, take a hexagon $h$ on the outside 
of $R$; assume $h$ is centered at $y \in N_x$. Then shift $R$ by $x-y$. 
Clearly, the hexagons on the outside of the shifted $C$-interior region are 
centered on the boundary sublattice. We must also check that the shifts do not create 
overlap.

To this end, let $h_1$ and $h_2$ be any two (distinct) hexagons in $C$-interior 
regions $R_1$ and $R_2$, and let $h_1'$ and $h_2'$ be the images of the hexagons 
under the shifts of $R_1$ and $R_2$ described in the preceding paragraph. 
If $R_1=R_2$ then clearly $h_1'$ and $h_2'$ do not overlap. Thus assume $R_1 \ne R_2$, 
and suppose $h_1$ and $h_2$ are centered at $y_1$ and $y_2$, respectively. Then $y_1 \in N_{x_1}$ 
and $y_2 \in N_{x_2}$ for some $x_1 \ne x_2$, and $h_1'$ and $h_2'$ 
are centered at $x_1$ and $x_2$, respectively. 
Since $x_1$ and $x_2$ are both points on the boundary sublattice, $h_1'$ and $h_2'$ 
do not overlap, as desired. \qed

\vs.1
Note that the configuration produced by the protocol in Lemma 3 does 
not necessarily produce a {\it legal} configuration, just a configuration 
with no overlaps.

\vs.2 \nd 
{\bf Lemma 4.}  The size of a contour $C$ is an integer.
\vs.1
\nd 
{\bf Proof.} Consider a configuration produced by the protocol in
Lemma 3. The new configuration has contours $C_1,C_2,\ldots ,C_m$
in place of the original contour $C$. Since the protocol creates no
overlaps, and since all the shifted $C$-interior regions remain within the region
enclosed by $C$, the contours $C_1,C_2,\ldots ,C_m$ have the same combined
area as the contour $C$. Furthermore, the shifted $C$-interior 
regions are all on the boundary sublattice, so the hexagons
bordering each $C_i$ are all on the
boundary sublattice. Thus we conclude that each $C_i$ could be completely
covered by nonoverlapping hexagons, all on the boundary sublattice. The
result follows. \qed

\vs.2
\nd 
{\bf Lemma 5.} Fix an outer contour $C$ of size $k$. There is a
one-to-one correspondence between legal configurations $A$ with
exactly $n$ interior hexagons and $C$ as an outer contour, and legal
configurations $A'$ with exactly $n+k$ interior hexagons.

\vs.1
\nd 
{\bf Proof.} Let $A_0$ and $A_1$ be two distinct configurations with
the outer contour $C$, and assume $A_0$ and $A_1$ have $n_0$ and $n_1$
interior hexagons, respectively. Using Lemmas 3 and 4, shift the $C$-interior
regions of $A_0$ and $A_1$ to produce 
configurations $\tilde A_0$ and $\tilde A_1$ which both have contours
$C_1,C_2,\ldots ,C_m$ that can be completely covered by nonoverlapping
hexagons. Cover these contours with nonoverlapping hexagons to produce
configurations $A_0'$ and $A_1'$ having $n_0+k$ and $n_1+k$ interior
hexagons, respectively. We claim first that $A_0'$ and $A_1'$ are legal
configurations; of course it suffices to show that $A_0'$ is a legal
configuration.

We have to show that each hexagon in $A_0'$ has a support. First
consider a hexagon $h$ in $A_0'$ in one of the shifted $C$-interior regions. 
Assume $h$ is on the inside of the shifted region. Because
shifts do not affect relative positions of hexagons inside the region,
and since the configuration was legal before the shift, $h$ must
have a support. Now assume $h$ is on the outside of the shifted
region. If $h$ does not have a support in the shifted region, then $h$
must have had a support outside the region before shifting. Since the
region $C_1\cup \ldots \cup C_m$ is completely filled with hexagons, one
of these must be a support of $h$. Finally consider a hexagon $h$
not in one of the shifted $C$-interior regions. Since the contours
$C_1,C_2,\ldots ,C_m$ were completely filled with hexagons, clearly $h$ has
a support.

Next, we claim that $A_0'$ and $A_1'$ are distinct
configurations. Note first that the $C$-interior regions of
$A_0$ and $A_1$ have identical outsides, because the contour $C$
defines these outsides. Thus there is an obvious pairwise
association between the $C$-interior regions of $A_0$ and the
$C$-interior regions of $A_1$. Since $A_0$ and $A_1$ are distinct,
either at least one of these pairs of $C$-interior regions, say $R_0$
and $R_1$, must have different insides, or $A_0$ and $A_1$
must be different outside the region enclosed by $C$. In the latter
case, the configurations $A_0'$ and $A_1'$ must be distinct, because
the shifts done by the protocol in Lemma 3 do not change anything
outside the region enclosed by $C$. In the former case, $R_0$ and
$R_1$ have distinct insides. This of
course does not change after shifting, and so $A_0'$ and $A_1'$ are
distinct. In either case $A_0'$ and $A_1'$ are distinct, so we have
the desired correspondence. \qed

\vs.2
\nd 
{\bf Lemma 6.} Let $C$ be a fixed contour of size $k$. The probability
that a configuration has the contour $C$ is at most $e^{-k\mu}$.

\vs.1
\nd 
{\bf Proof.} To prove this, we use the association in Lemma 5. Let $Z$
be the normalization, and let $E_C$ be the event
that a configuration has the contour $C$. Let $H_n$ be the number of
legal configurations $A$ having the contour $C$ and $n$ interior
hexagons, and let $H_n'$ be the number of legal configurations $A'$
having $n+k$ interior hexagons. By Lemma 5 we have that $H_n' \ge H_n$,
and of course we also have that $\sum_{n=0}^\infty e^{(n+k)\mu} H_n'
\le Z$. Thus, we have the estimate

\vs.1

$$Pr(E_C)= {{1}\over{Z}}\sum_{n=0}^\infty e^{n\mu} H_n \le
{{\sum_{n=0}^\infty e^{n\mu} H_n}\over{\sum_{n=0}^\infty e^{(n+k)\mu}
H_n'}} \le e^{-k\mu}$$

\vs.1
\nd
as desired. Note that this estimate is independent of the size of the
container $V$. \qed
\vs.2

We are finished with half of the Peierls argument. Now we provide
an upper bound on the number of contours of a given size. We do this
by counting graphs whose vertices are triangles in a
contour. Note that there are $6s^2$
triangles inside a hexagon.

Before we begin the counting argument we need the following
well-known facts from graph theory:

\vs.2
\nd 
{\bf Lemma 7\ a.} Let $T$ be a spanning tree for a set of $n$
points. Then $T$ has $n-1$ edges.
\vs0 \nd 
\hs.8 {\bf b.} A graph $G'$ produced by duplicating every edge of a
graph $G$ is Eulerian.

\vs.2

Now to count the contours, we make the following observation about the
structure of a contour $C$. The union of all the 
triangles in a contour consists of several disjoint connected components. 
These components are joined to neighboring components by line segments in 
$C$ of length strictly less than $s$; recall that such line segments are 
the intersections of neighboring hexagons. Thus, 
the minimum number of lattice segments in a
path between triangles in neighboring components is at most
$s-1$, where by a {\it lattice segment} we mean a line segment joining 
nearest neighbor lattice sites. This leads to the following lemma.

\vs.2
\nd 
{\bf Lemma 8.} Suppose $C$ is a contour of size $k$. Then $m=6s^2k$ is
the number of triangles in the contour. There is a sequence
$(t_1,\dots ,t_{2m-1})$ of triangles in $C$ such that each triangle 
in $C$ is some $t_i$, and such that the mimimum number of
lattice segments in a path joining $t_i$ and $t_{i+1}$ is $\le s-1$.

\vs.1
\nd 
{\bf Proof.} Let $Z$ be a set of vertices, one for each triangle
in $C$. Define the distance between vertices in $Z$ as one plus the
minimum number of lattice segments in a path joining the corresponding
triangles in $C$. Partition $Z$ into $Z_1,\ldots ,Z_k$, where the
$Z_i$ correspond to the connected components of the union of all the
triangles in $C$. For each $i$, join two vertices in $Z_i$ by an
edge iff the distance between them is $1$. Then one by one remove
edges comprising cycles in each $Z_i$ (this process is not necessarily
unique).

Next, for each $i \ne j$, join $x \in Z_i$ to $y \in Z_j$ by an edge
iff the distance between $x$ and $y$ is less than or equal to $s$. By
preceding considerations we see that the resulting graph is
connected. One by one remove edges comprising cycles to produce a tree
$T$ spanning all the vertices of $Z$. Note that all $(m-1)$ of the
edges of $T$ have length $\le s$.

Now define a duplicate graph $D$ which has the same vertices as $T$
but which has two edges joining each pair of vertices which are joined
by an edge in $T$. Then $D$ is an Eulerian graph, so there is an
Eulerian path, that is, a path $\Gamma$ in $D$ which traverses every
edge exactly once. $D$ has $2(m-1)$ edges, so $\Gamma$ traverses 
$2(m-1)+1$ vertices, counting repeats. Clearly $\Gamma$ traverses
each vertex of $T$ at least once. So take $t_i$ to be the triangle
 corresponding to the $i$th vertex traversed by $\Gamma$. \qed

\vs.2
\nd 
{\bf Lemma 9.} The number of contours $C$ of size $k$ such that the 
origin is in the region enclosed by $C$ is less than $p_m q^m$, 
where $m=6s^2k$, $p_m=6m^2$ and $q=36(s+1)^4$.

\vs.1
\nd 
{\bf Proof.} Using Lemma 8, for any contour $C$ of size $k$ we have a
corresponding sequence $(t_1,\ldots t_{2m-1})$ of triangles in $C$,
where $m=6s^2k$. Moreover, since the sequence covers all the triangles 
in $C$, distinct contours are associated with distinct
sequences (note that a contour is totally defined by the positioning
of its triangles). There are no more than $6m^2$ possible triangles
 that a contour enclosing the origin can contain, and given the position of the
$i$th triangle there are at most $6(s+1)^2$ possibilities for the
position of the $(i+1)$st. Since there are $2m-1<2m$ total elements of
the sequence, we may take $q=36(s+1)^4$ and $p_m=6m^2$ to get the
desired result. \qed

\vs.2

Now we are ready to combine the two main ingredients of the Peierls
argument into the final result:

\vs.2
\nd 
{\bf Theorem 1.} The probability that there is a hexagon centered 
in $S$ which is on the boundary sublattice goes to $1$ as $\mu$ goes to
infinity. That is, $Pr(0_{B})\to 1$ as $\mu \to \infty$.

\vs.1
\nd 
{\bf Proof.} By Corollary 1, Lemma 6 and Lemma 9,
$Pr([0_{B}]^c)$ is bounded above by
$\sum_{k=1}^\infty p_m(q^{6s^2}e^{-\mu})^k$, where again $m=6s^2k$. Since
$q^{6s^2}$ is a constant depending only on $s$, and since $p_m$ is
polynomial in $k$, we have that for $\mu$ sufficiently large, the summation
bounding $Pr([0_B]^c)$ is arbitrarily small. 
Note that the estimate underlying this result is independent of the 
size of the container $V$. \qed

\vs.2

We have abbreviated this result by saying that at sufficiently large $\mu$
the system has long range order. We also have the following percolation result.

\vs.2 \nd {\bf Corollary 2.} The probability that the origin lies inside 
a hexagon linked to the boundary goes to $1$ as $\mu$ goes to
infinity; that is, $Pr(0_{LB}) \to 1$ as $\mu \to \infty$.  

\vs.1 \nd
{\bf Proof.} This follows by Lemma 2 and the same argument as in
Theorem 1. \qed

\vs.3
\nd
{\bf 2. Numerical Results for low $\mu$.}
\vs.2

We ran Markov chain Monte Carlo simulations on the model for a range
of values of $\mu$. We checked that the Monte Carlo runs were not
sensitive to the initial condition (see Figure 4); since lower volume
fraction initial conditions tended to equilibriate faster, we started
the remaining Monte Carlo runs with void configurations. 

If $\mu << 0$ typical configurations do not fill the container -- see
Figures 5 and 6 -- and it is harder to develop useful data. 
Our goal in this section is to show that in the infinite volume limit the boundary
has no influence near the origin for small $\mu$. 
As the main object of our simulation we consider the quantity $p(\mu)$, 
defined as follows.  
First recall the set $S$ defined in the preceding section, namely 
a set of representative lattice sites for each of the $3s^2$ different sublattices, such 
that $S$ is contained inside a regular hexagon of side length $s$ centered at 
the origin. Recall that we define a {\it sublattice} to be a set of lattice sites 
corresponding to the centers of a collection of hexagons which tile the plane.
\vs.05 \nd
{\bf Definition 2.} 
For a fixed container $V$, we define $p(\mu)$ to be the 
probability that there is a hexagon $h$ centered in $S$ such that $h$ is 
centered in the boundary sublattice.
\vs.05 \nd
Note that $p(\mu)$ is the same as the quantity $P(0_B)$,
but here we emphasize its dependence on $\mu$.
We want to show that in the infinite volume limit, 
$p(\mu)$ is constant in some interval of positive length; 
its value there should be $1/3s^2$, where $3s^2$ is the number 
of sublattices. Recall from our results in the previous section 
that $p(\mu) \to 1$ as $\mu \to \infty$, {\it uniformly in system size}.

Our argument will concentrate on the interval $[1,2]$ for $\mu$. 
Simulation for $\mu$ inside this interval and inside the interval 
$[0,10]$ suggests that in the infinite volume limit, 
$p(\mu)$ is indeed constant inside $[1,2]$ (see Figures 7-12). 

To obtain numerical estimates of $p(\mu)$
we considered the following functions on our Monte Carlo runs. 
For a configuration $A$ we let $\delta(A) = 1$ if there is a 
hexagon in $A$ centered at a point in $S$ in the boundary sublattice; we let 
$\delta(A) = 0$ otherwise. We define $t(A) = 1$ if there is a hexagon in $A$ 
centered at a point in $S$, and $t(A) = 0$ otherwise.

For systems ranging in volume from 
$276$ to $1151$ (in units of hexagon volume) we evaluated
$\delta$ and $t$ on configurations $A_1,A_2,\ldots$, and
for each system size we consider the following statistic:

\vs.2 \nd 
$$p_M := {1\over T}[\delta(A_1)+\delta(A_2)+\cdots +\delta(A_M)]$$
\vs.1 \nd
where
\vs.1 \nd
$$T=t(A_1)+t(A_2)+\cdots +t(A_M)$$
\vs.2

So the expected value of $p_M$ is exactly $p(\mu)$.
 We obtain confidence intervals for $p_M$ in the same way as in [AR]; 
in particular we determine the mixing time for our simulations
using the biased autocorrelation function on volume fraction data, and then 
 use the common method of batch means [Ge] with about $10$ batches 
for each run, with batch size $M$ chosen so that there are at least $5$ 
mixing times per batch (except in the transition region).

If the boundary has no influence near the origin, hexagons 
should appear in each sublattice with equal probability, so 
we expect that the limiting value of $p(\mu)$ is 
exactly $1/(3s^2)$ for small $\mu$.
In Figures 7-12 we compare data from our Monte
Carlo runs to the line $y = 1/(3s^2)$, with $s=3$.  
For $\mu \in [0,4]$ the data suggests that
$p(\mu)$ follows the line; then in the range $\mu \in [4,6]$, 
$p(\mu)$ increases to about $1$; for $\mu > 6$, $p(\mu)$ stays near the 
line $y = 1$. In Figures 11-12 we consider more detailed data for $\mu$ in $[1,2]$. 
Our $95\%$ confidence intervals cover the line $y = 1/(3s^2)$ more
than $95\%$ of the time, as appropriate.

We note that the transition region changes as $s$ increases. In
particular, as $s$ increases the smallest value of $\mu$ such 
that $p(\mu) > 1/(3s^2)$ seems also to increase; compare Figures 9, 13, 14. 

\vs.3 

\nd {\bf 3. Conclusion.} \vs.1  

Our argument is based on the behavior of $p(\mu)$ -- the
probability that a hexagon near the origin is on the same sublattice
as the boundary hexagons -- as a function of the parameter $\mu$, the
variable controlling average volume fraction.
We have proven that for sufficiently large positive $\mu$,
 $p(\mu)$ is greater than $1/(3s^2)$, and in fact $p(\mu)$ 
approaches 1 as $\mu\to \infty$, uniformly in the size of the system.
 In addition we have numerical evidence that in
an interval above zero, $p(\mu)$ has the constant value $1/(3s^2)$ 
in the infinite volume limit,
indicative of disorder. As the two types of behavior 
cannot be connected analytically we conclude [FR] that
the model undergoes a phase transition at some
positive $\mu$. The transition can be seen in Figure 9, but it would
take much more simulation to demonstrate singular behavior at a
specific value of $\mu$. Instead, our
argument for the existence of a transition is based on failure of
analyticity. (To use simulation to show that {\it dependence} on the
boundary survives in the infinite volume limit
requires careful study of the size of the simulation samples, while
the burden is easier to show {\it independence} of the boundary, as we do.)

The transition we have found is of the order/disorder
 type since the long range order which we prove to hold at large
 $\mu$ is absent at low $\mu$. We note that, as usual in hard-core lattice models
[HP], our results only apply for a finite ratio $s$ of hexagon size to
lattice spacing; our upper bound on ordered behavior diverges as $s \to \infty$.

\vfill \eject


\hbox{}
\vs-.3
\centerline{\bf Simulation Results}
\vs0 
\hbox{}
\vs0 \nd
\epsfig .9\hsize;  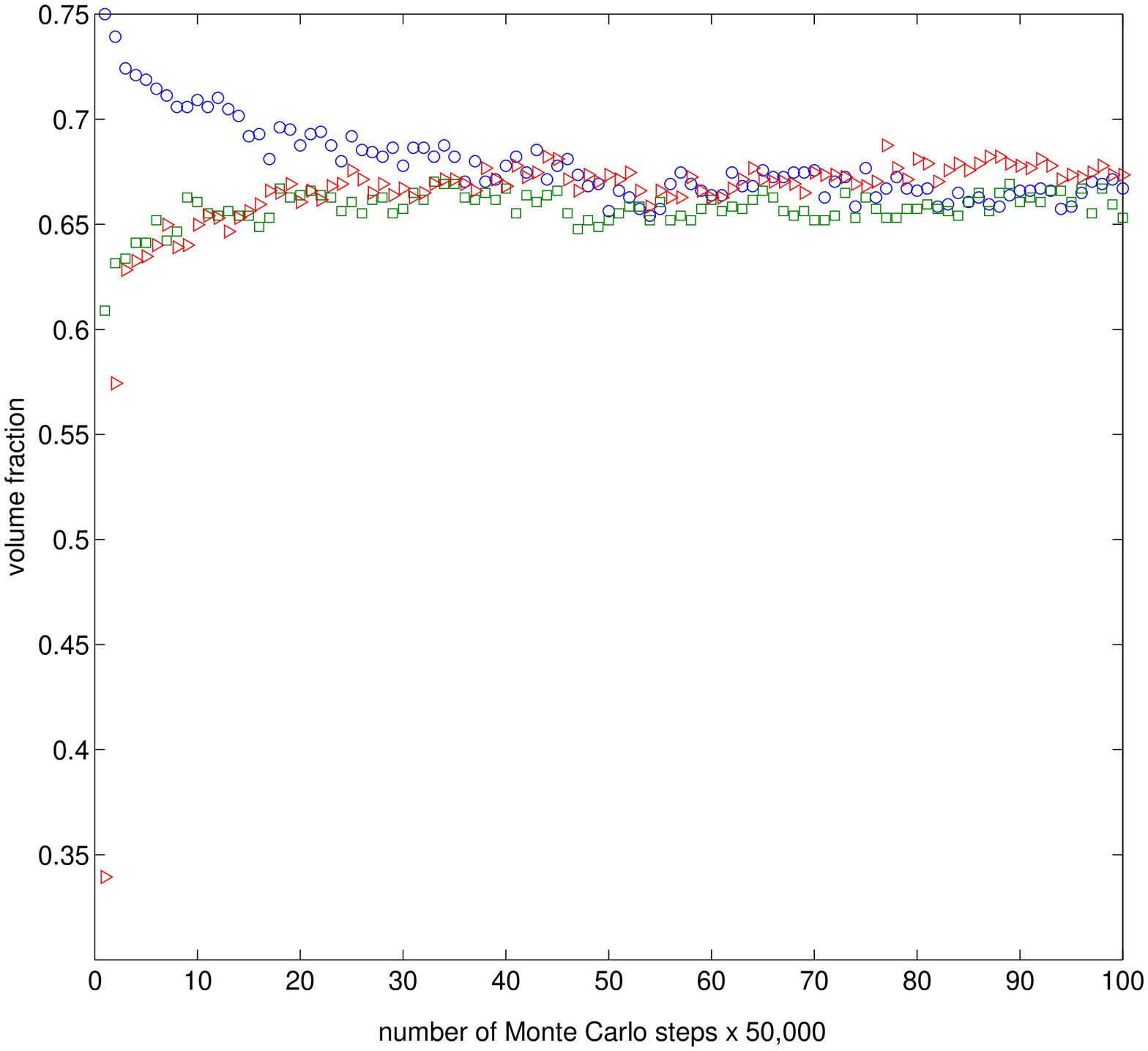;
\vs-.1 \nd
Figure 4. Plot of volume fraction versus number of moves, 
from three different initial volume fractions, for a system of volume
$729$ and $\mu=1$.

\vfill \eject
\hbox{}
\vs0 \nd
\epsfig .9\hsize; 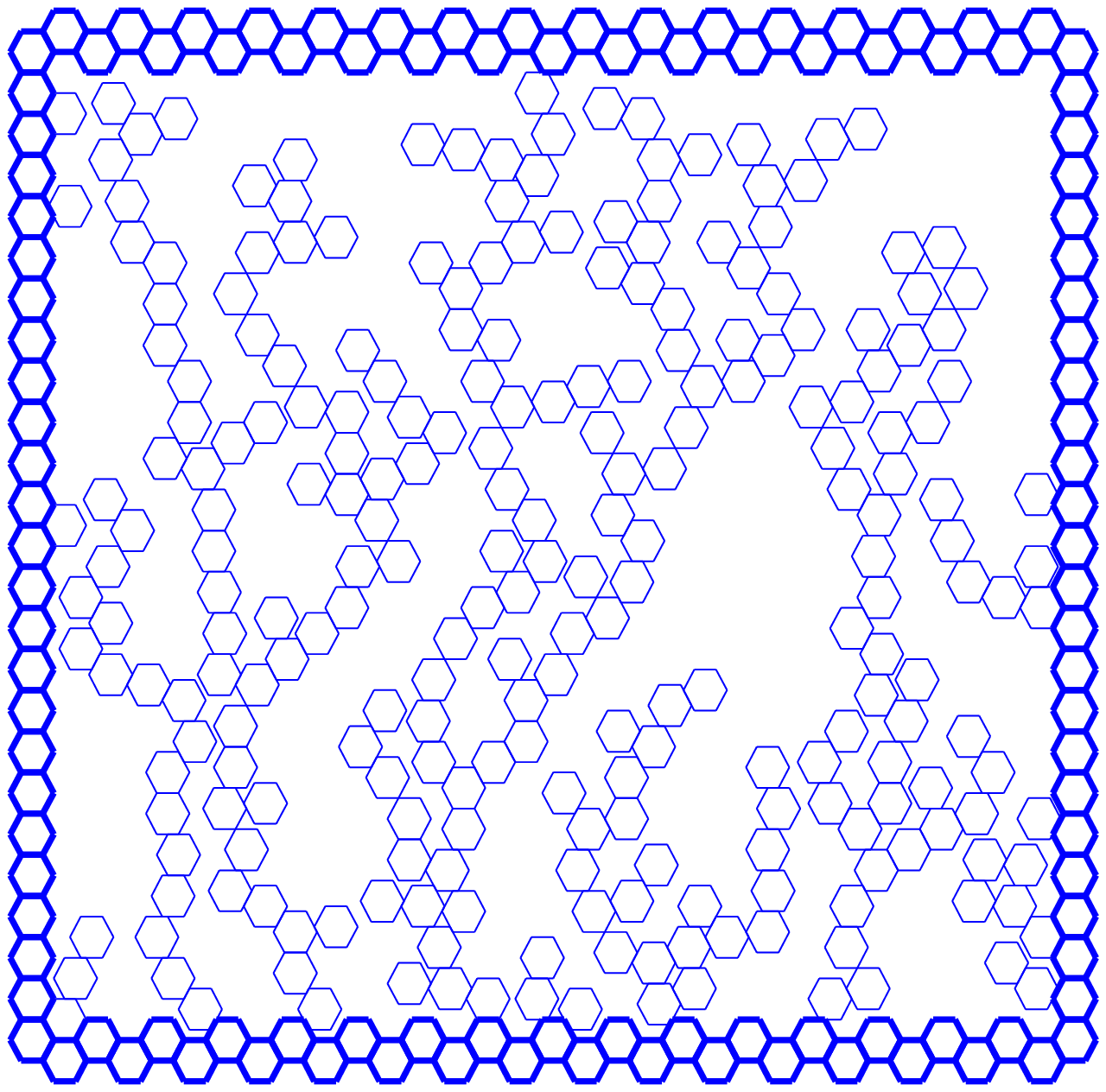;
\vs0 \nd
Figure 5. A configuration of $250$ hexagons in equilibrium at $\mu = -4$, in a 
system of volume $729$.

\vfill \eject
\hbox{}
\vs0 \nd
\epsfig .9\hsize; 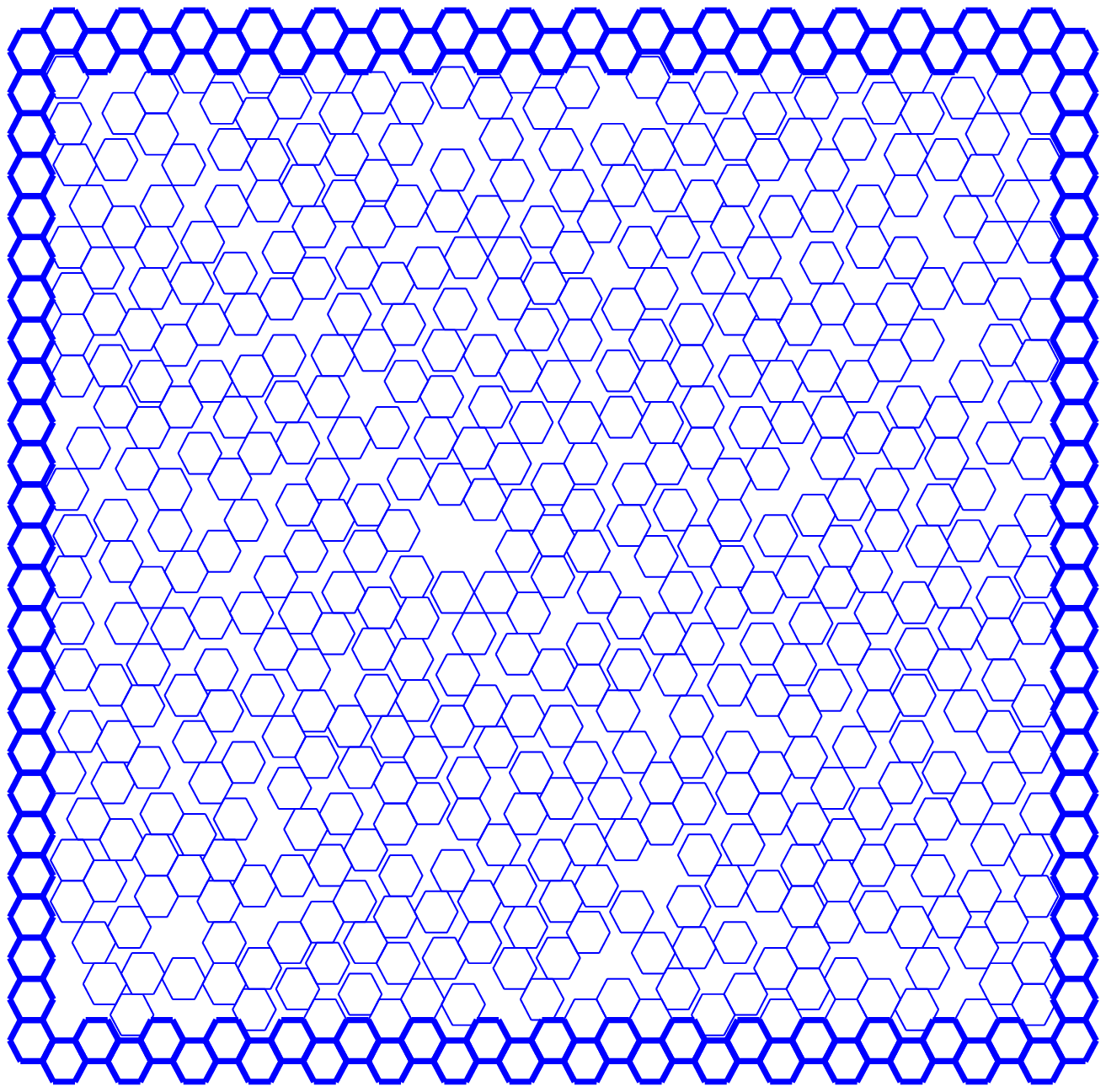;
\vs0 \nd
\centerline{Figure 6. A plot of a configuration in equilibrium at $\mu = 1$}.

\vfill \eject
\hbox{}
\vs0 \nd
\epsfig .9\hsize; 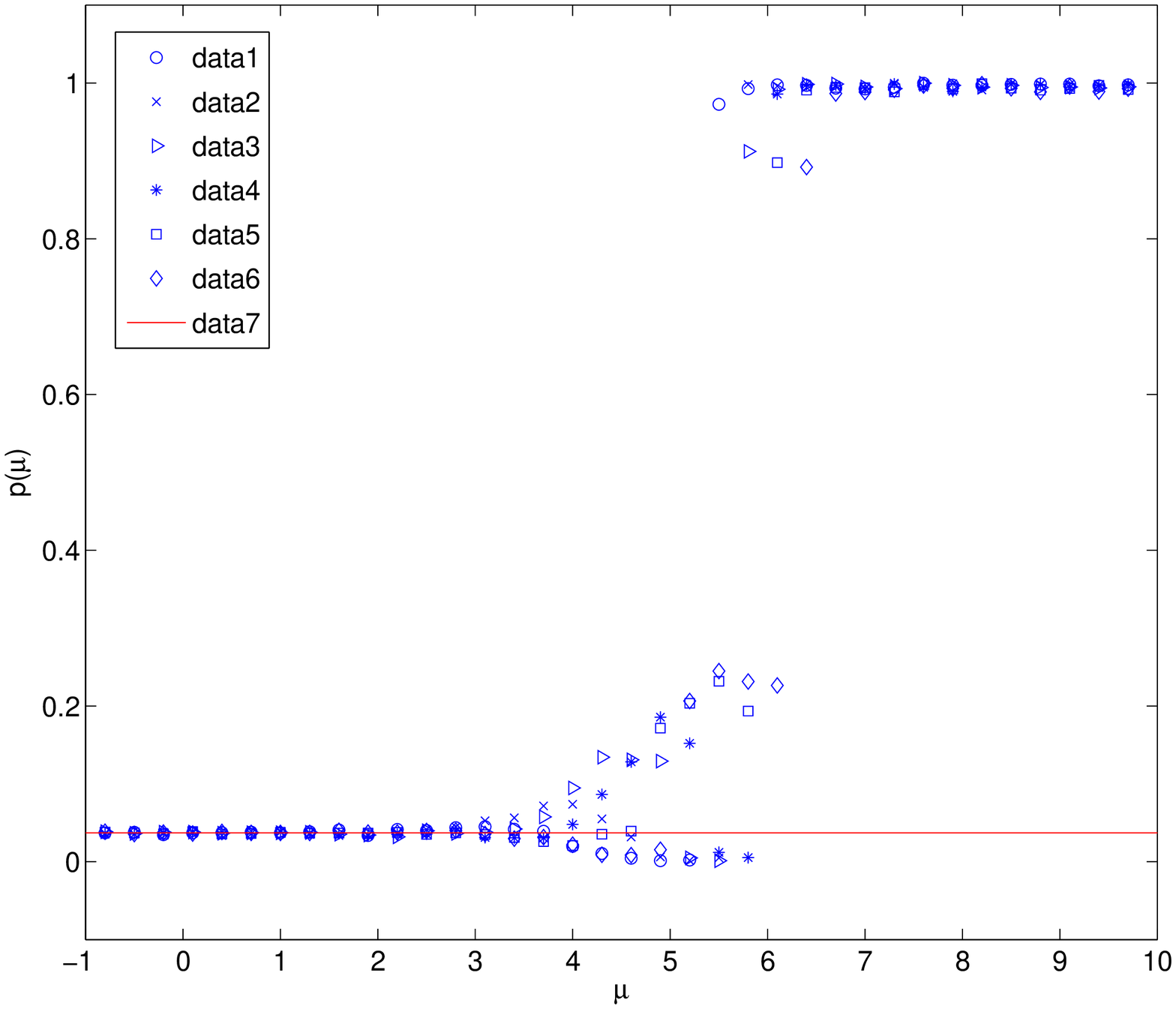;
\vs0 \nd
Figure 7. Plot of $p(\mu)$ vs.\ $\mu$ for systems of volume $276$ (data1) to 
$1151$ (data6), for $s=3$. Data7 is the line $p(\mu) =
{{1}\over{3s^2}}={1\over 27}$.

\vfill \eject
\hbox{}
\vs0 \nd
\epsfig .9\hsize; 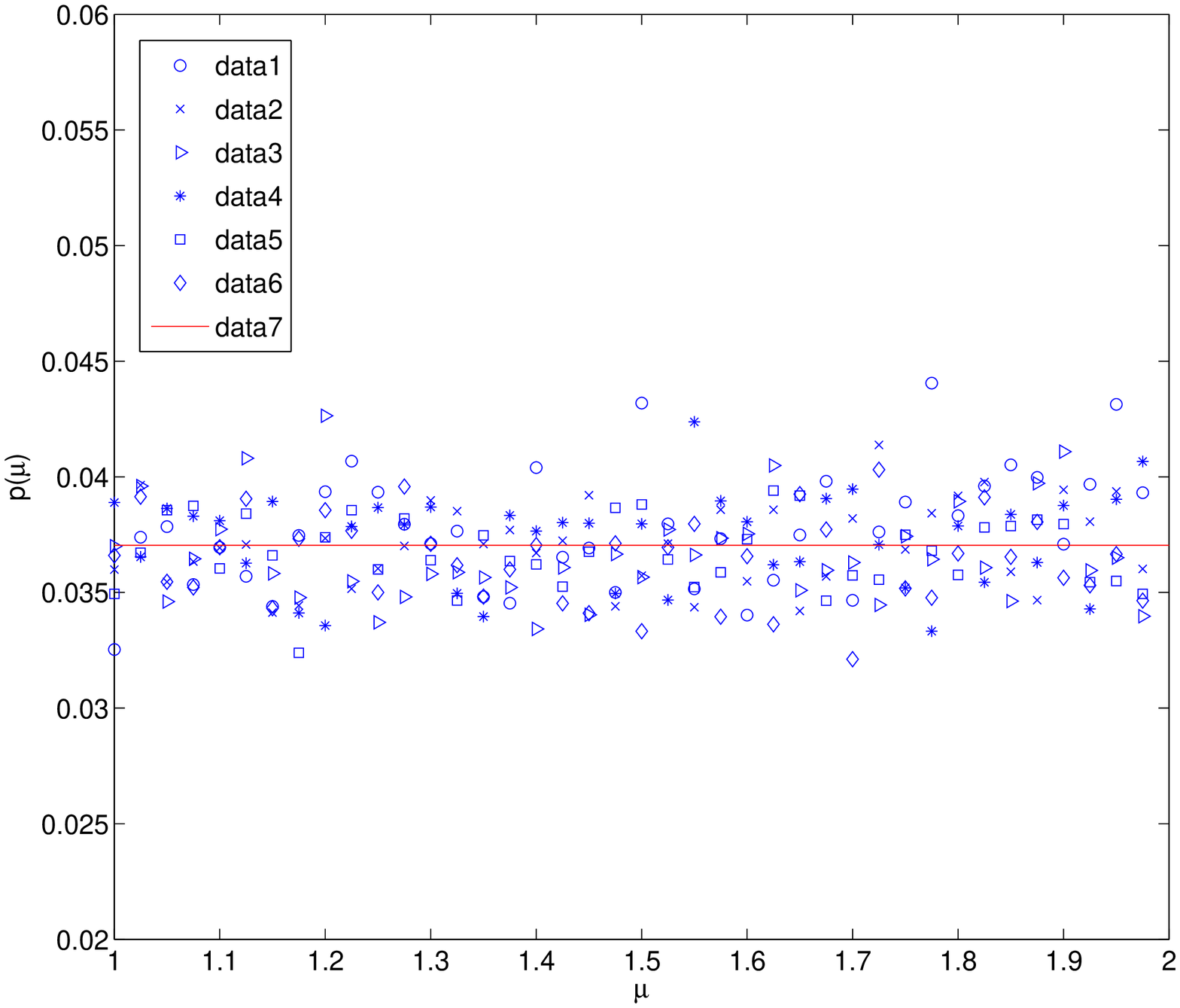;
\vs0 \nd
Figure 8. Plot of $p(\mu)$ vs.\ $\mu$ for systems of volume $276$ (data1) to 
$1151$ (data6), for $s=3$. Data7 is the line $p(\mu) = {{1}\over{3s^2}}={1\over 27}$.

\vfill \eject
\hbox{}
\vs0 \nd
\epsfig .9\hsize; 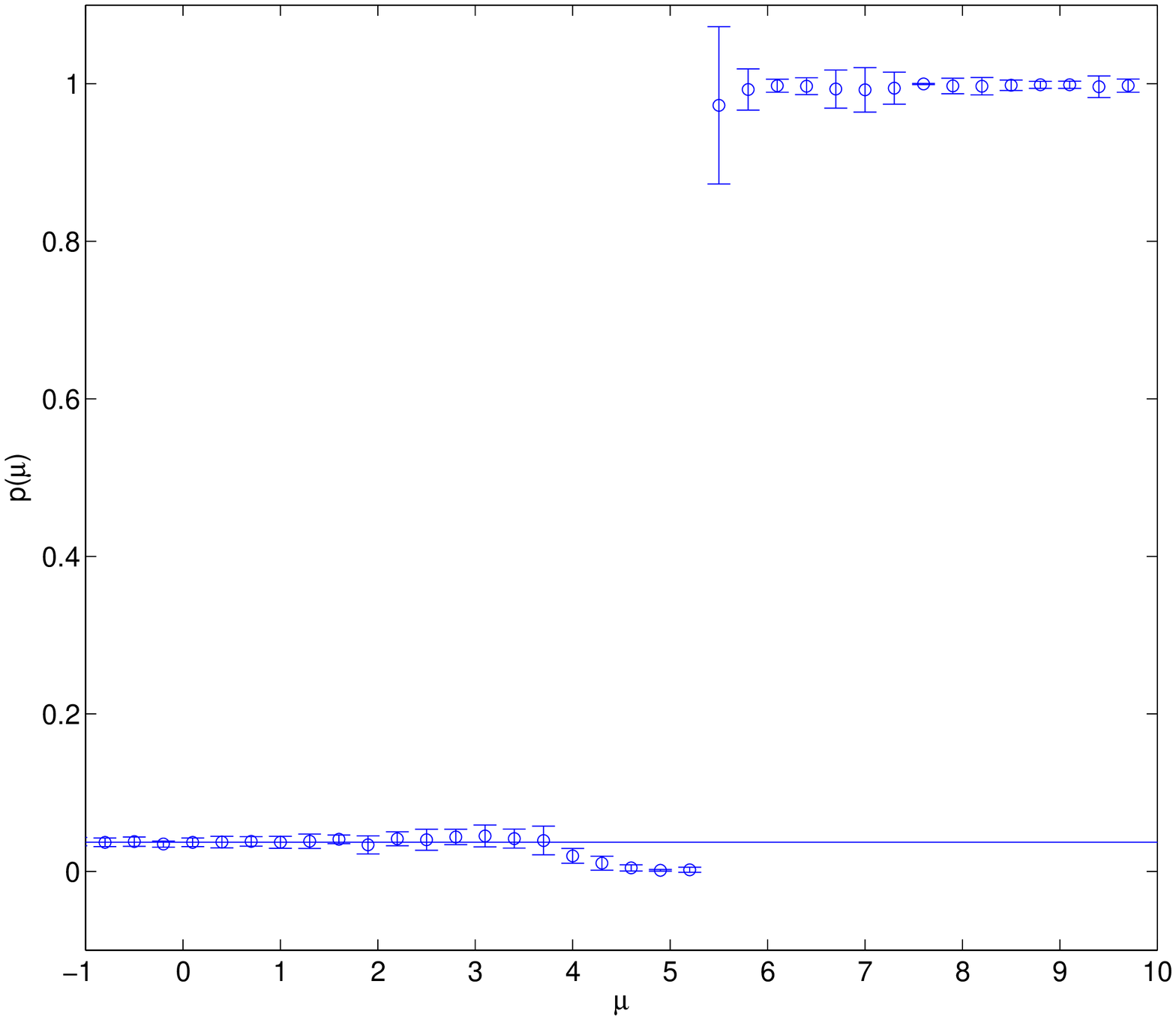;
\vs0 \nd
{Figure 9. Plot of $p(\mu)$ vs.\ $\mu$ for a system of volume $276$,
with error bars, for $s=3$}. The line is $p(\mu)={{1}\over{3s^2}}={1\over 27}$.

\vfill \eject
\hbox{}
\vs0 \nd
\epsfig .9\hsize; 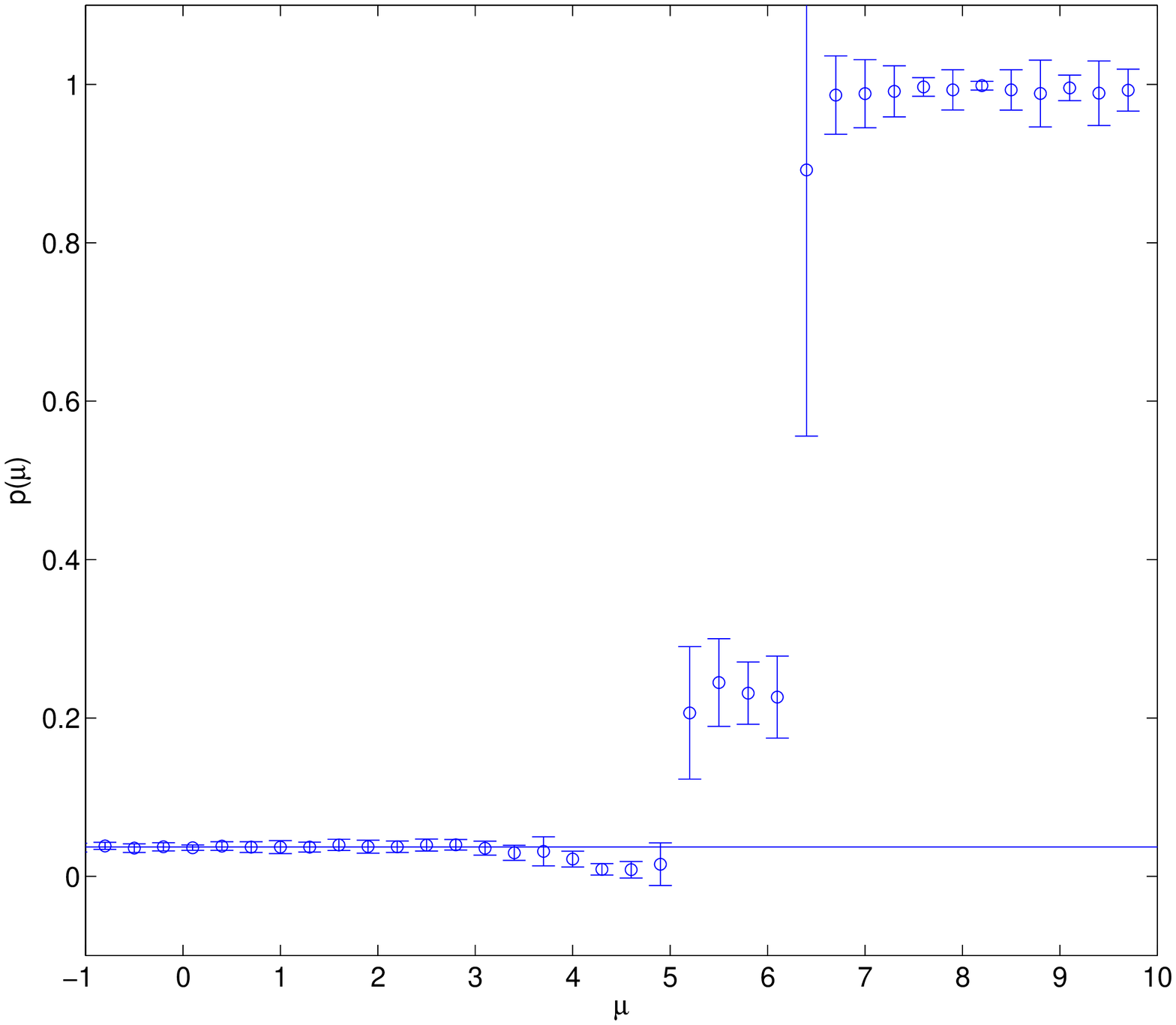;
\vs0 \nd
{Figure 10. Plot of $p(\mu)$ vs.\ $\mu$ for a system of volume $1151$,
with error bars, for $s=3$}. The line is $p(\mu)={{1}\over{3s^2}}={1\over 27}$.

\vfill \eject
\hbox{}
\vs0 \nd
\epsfig .9\hsize; 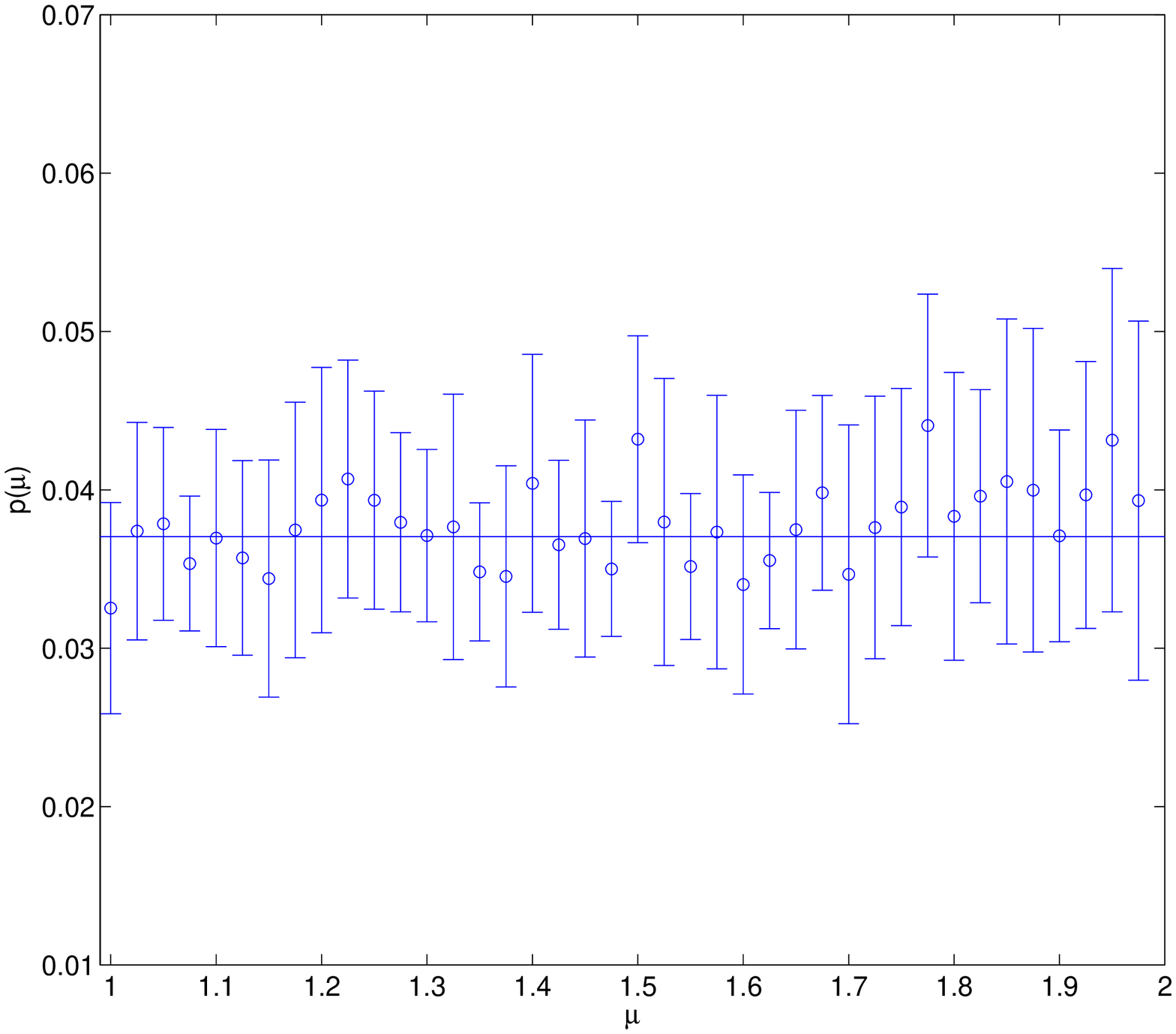;
\vs0 \nd
{Figure 11. Plot of $p(\mu)$ vs.\ $\mu$ for a system of volume $276$,
with error bars, for $s=3$}. The line is $p(\mu)={{1}\over{3s^2}}={1\over 27}$.

\vfill \eject
\hbox{}
\vs0 \nd
\epsfig .9\hsize; 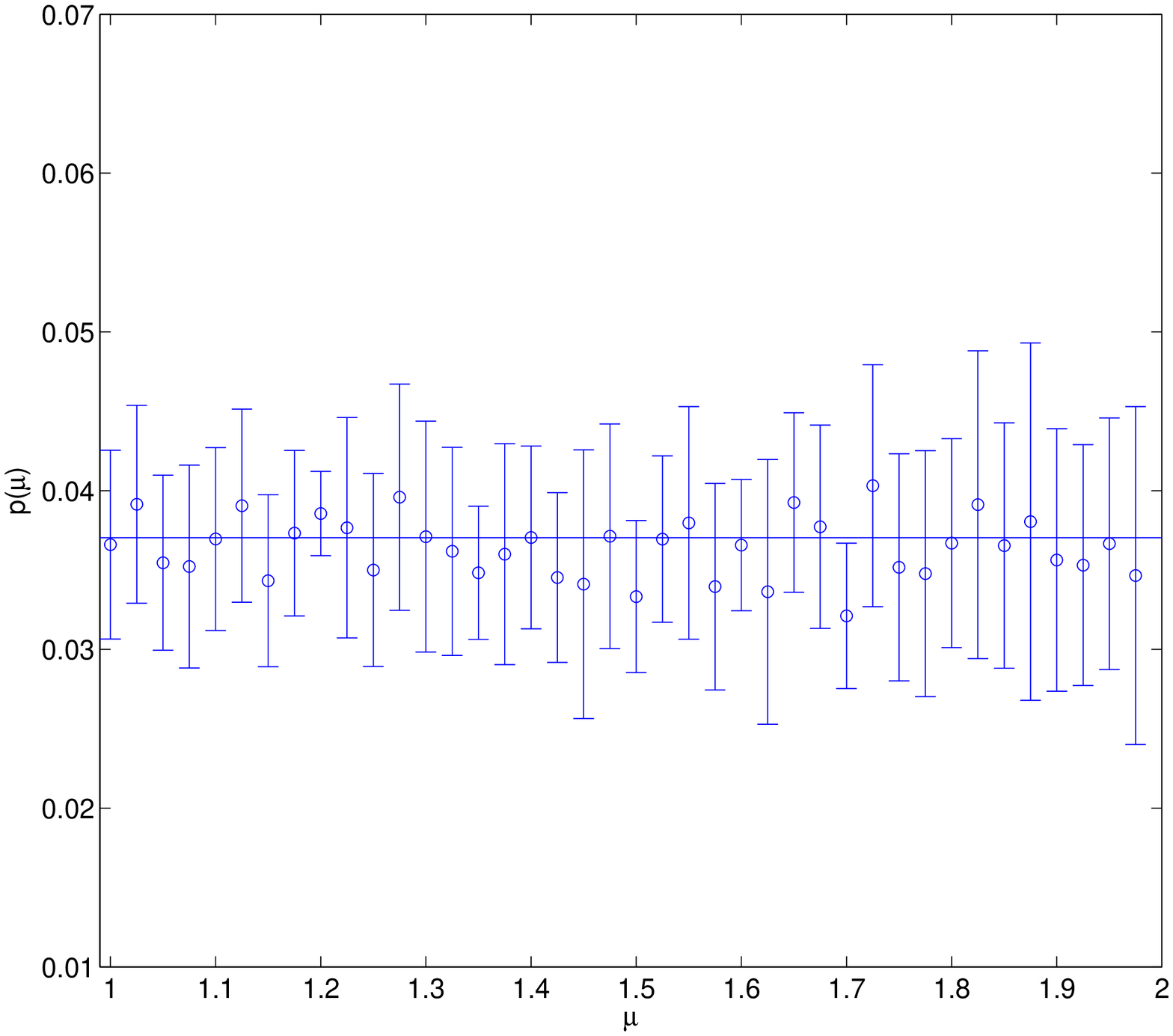;
\vs0 \nd
{Figure 12. Plot of $p(\mu)$ vs.\ $\mu$ for a system of
volume $1151$, with error bars, for $s=3$}. The line is 
$p(\mu)={{1}\over{3s^2}}={1\over 27}$. 

\vfill \eject
\hbox{}
\vs0 \nd
\epsfig .9\hsize; 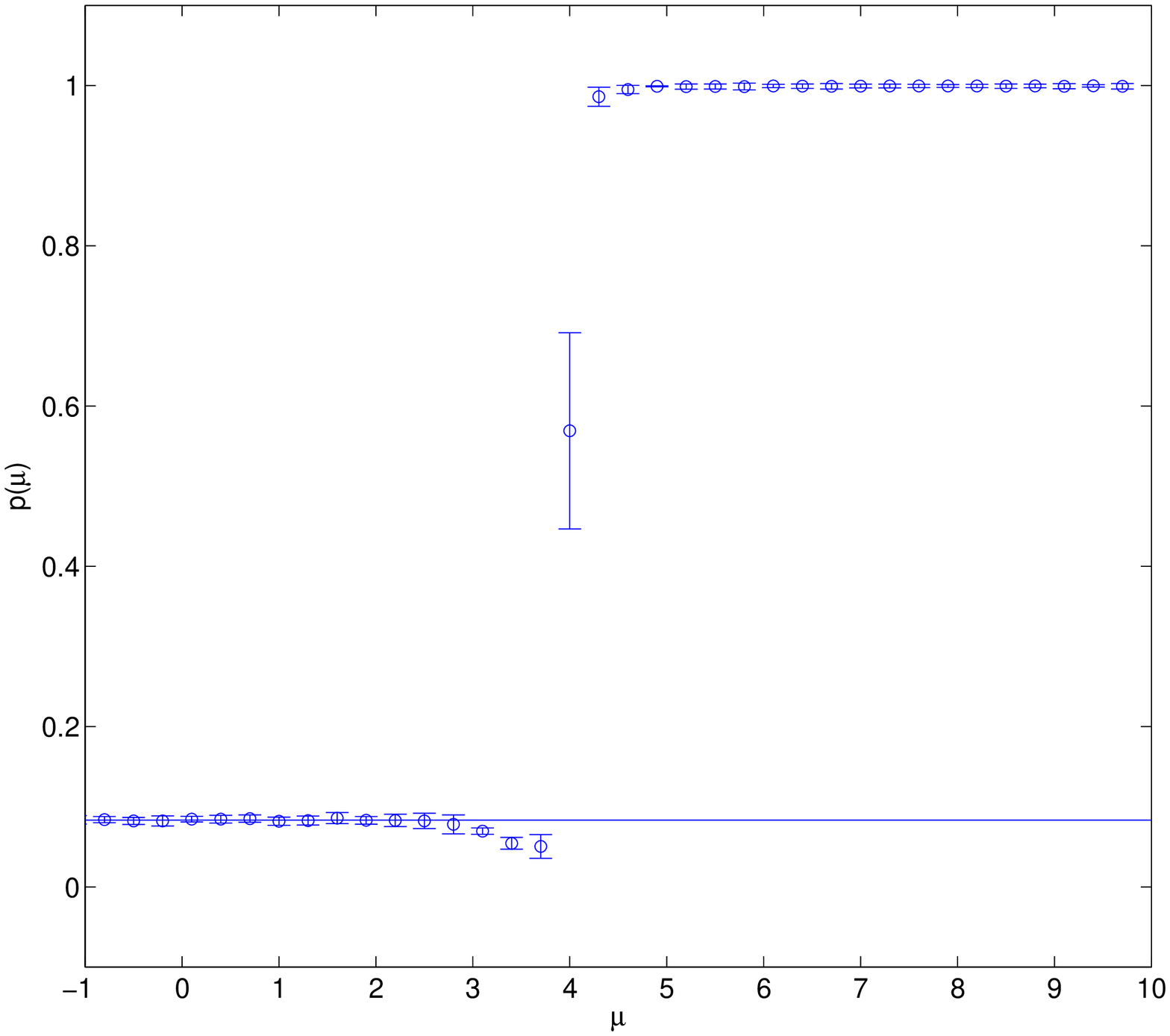;
\vs0 \nd
{Figure 13. Plot of $p(\mu)$ vs.\ $\mu$ for a system of volume $276$,
for $s=2$}. The line is $p(\mu)={{1}\over{3s^2}}={1\over 12}$. 


\vfill \eject
\hbox{}
\vs0 \nd
\epsfig .9\hsize; 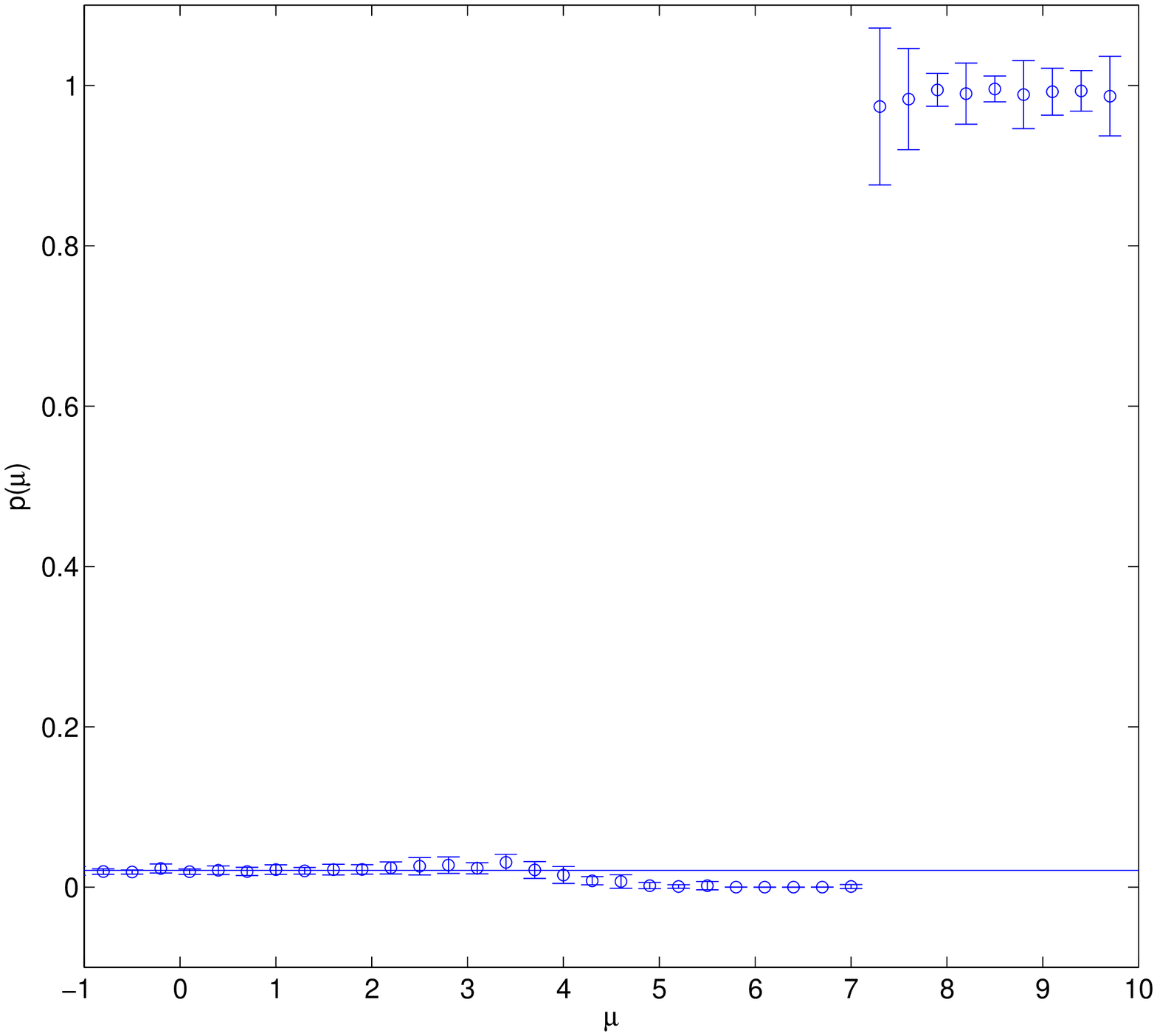;
\vs0 \nd
{Figure 14. Plot of $p(\mu)$ vs.\ $\mu$ for a system of volume $276$,
for $s=4$}. The line is $p(\mu)={{1}\over{3s^2}}={1\over 48}$. 

\vfill \eject
\vs.2
\centerline{\bf References}
\vs.2
\item{[AR]} D. Aristoff and C. Radin, Random loose packing in granular matter, 
J. Stat. Phys., 135(2009), 1-23. 
\item{[dG]} P.G. de Gennes, Granular matter: a tentative view. Rev. Mod.
  Phys. {71} (1999) S374--S382.
\item{[EO]} S.F. Edwards and
  R.B.S. Oakeshott, Theory of powders, Physica A 157 (1989) 1080-1090.
\item{[FR]} M.E. Fisher and C. Radin, Definitions of thermodynamic 
  phases and phase transitions, workshop report,
  \vs0 
  http://www.aimath.org/WWN/phasetransition/Defs16.pdf
\item{[Ge]} C.J. Geyer, Practical Markov chain Monte Carlo,
Stat. Sci. 7 (1992) 473-483.
\item{[Gi]} J. Ginibre, On some recent work of Dobrushin,
{\it Syst\`emes \`a un nombre infini de degr\'es de libert\'e},
CNRS, Paris, 1969, pp. 163-175.
\item{[HP]} O. J. Heilmann and E. Praestgaard, Phase transition of hard hexagons
on a triangular lattice,
J. Stat. Phys. 9 23-44 (1973).
\item{[LY]} T. D. Lee and C. N. Yang, Statistical Theory of Equations 
of State and Phase Transitions. II. Lattice Gas and Ising Model,
Phys. Rev. 87 410-419 (1952).
\item{[MP]} R. Monasson and O. Pouliquen, 
Entropy of particle packings: an illustration on a toy model,
{Physica A} 236 (1997) 395-410.
\item{[Ra]} C. Radin, Random close packing of granular matter, 
J. Stat. Phys. 131 (2008) 567-573.
\item{[Ru]} D. Ruelle, {\it Statistical Mechanics; Rigorous Results}, Benjamin, New
York, 1969.

\end